\documentclass[aps,pra,twocolumn,showpacs]{revtex4}

\usepackage{bm}

\begin{document}

\title{Continuous-variable Werner state:
separability, nonlocality, squeezing and teleportation}

\author{Ladislav Mi\v{s}ta Jr., Radim Filip, and Jarom\'{\i}r Fiur\'{a}\v{s}ek}
\affiliation{Department of Optics, Palack\'y University,
17. listopadu 50, 772 00 Olomouc, Czech Republic}

\date{\today}

\begin{abstract}
We investigate the separability, nonlocality and squeezing of
continuous-variable analogue of the Werner state: a mixture of pure two-mode
squeezed vacuum state with local thermal radiations. Utilizing this
Werner state, coherent-state teleportation in Braunstein-Kimble setup is
discussed.
\end{abstract}
\pacs{03.65.Bz}

\maketitle

\section{Introduction}

Quantum entanglement and nonlocality are fundamental
resources for the quantum information processing such as
quantum teleportation \cite{Bennett93}, entanglement swapping
\cite{Pan98}, dense coding \cite{Bennett92},
quantum cryptography \cite{Ekert91} and quantum computation \cite{Barenco95}.
The efficiency of quantum information processing significantly depends on
the degree of entanglement or nonlocality of the quantum state
shared by the parties involved in a given protocol.
This dependence may be particularly vividly illustrated
with the Werner states \cite{Werner89}, which are formed by a
mixture of maximally entangled state and a separable maximally mixed
state,
\begin{equation}\label{werner}
{\rho}=p|\Psi\rangle\langle\Psi|+\frac{(1-p)}{d^2}{I}_1\otimes{I}_2,
\quad 0\leq p\leq 1,
\end{equation}
where
\begin{equation}\label{maxim}
|\Psi\rangle=\frac{1}{\sqrt{d}}\sum_{i=1}^{d}|i\rangle_1|i\rangle_2
\end{equation}
is maximally entangled state in $d$-dimensional Hilbert space.

The Werner state is characterized by a single  parameter:
the probability $p$ of the maximally entangled state in the mixture
and the Werner state is entangled iff $p>1/(1+d)$. When the Werner state
is used as a quantum channel for teleportation, then the average teleportation
fidelity is given by $F=p+(1-p)/d$ \cite{Horodecki99b, Badziag00}. This figure
should be compared with the maximum fidelity achievable by means of classical
communication and local operations $F_C=2/(1+d)$ \cite{Massar95}.
Since this boundary is reached exactly for $p=1/(1+d)$ one concludes that
all entangled Werner states are useful for the teleportation. Particularly
interesting is the Werner state of two qubits, because for this system
both the necessary and sufficient conditions on inseparability and
nonlocality have been  established by Horodeckis
\cite{Horodecki95,Peres96,Horodecki96}. An important feature of the
two-qubit Werner states is a non-empty gap $1/3<p\leq 1/\sqrt{2}$ between
separable and nonlocal states.

In recent years, great attention has been paid to the quantum
information processing with continuous variables (CV).
Most protocols developed originally for discrete quantum variables (qubits)
have been  extended to continuous variables, namely
teleportation \cite{Braunstein98},
dense coding \cite{Braunstein00}, entanglement swapping \cite{Polkinghorne99}
and quantum cloning \cite{Cerf00}.

In this paper, we introduce a natural analogue of the Werner state
(\ref{werner}) for the CV systems:
a mixture of pure two-mode squeezed vacuum state
and mixed separable thermal state. We analyze in detail the
separability, nonlocality and squeezing of the CV Werner state and we also discuss its
usefulness in the teleportation of coherent states.
Since there is no general method how to test the separability of a generic
state in infinite-dimensional Hilbert space, one has to resort to
some particular tests. We use the Peres-Horodecki (PH) criterion based on
partial transposition \cite{Peres96}. Remarkably, non-positive partial
transpose is necessary and sufficient condition for inseparability of
two-mode bipartite Gaussian states \cite{Duan00}. However, the CV Werner state
discussed in this paper is not Gaussian and hence in our case the PH criterion
provides only a sufficient condition on the entanglement.

Testing of nonlocality for CV systems is based predominantly on the
Banaszek-W\'{o}dkiewicz form of the Bell inequalities \cite{Banaszek98}
working with Wigner function of a state. Here, we suggest alternative CHSH
Bell inequalities for continuous quantum variables.
By means of specific local transformations we map the two-mode CV Werner state
onto state of two qubits and then we employ the necessary and sufficient
conditions on nonlocality for two-qubit system.

After discussing the separability and nonlocality of the Werner state
we analyze its performance in quantum information processing.
We consider the standard Braunstein-Kimble (BK) scheme for teleportation
of CV \cite{Braunstein98} where the Werner state serves as a quantum channel.
Specifically, we focus on the teleportation of coherent states and we compare
our findings with the results that have been obtained for qubit or qudit
teleportation with Werner states \cite{Horodecki99b, Badziag00}.

The paper is organized as follows. In Sec.~II, the CV analogue of Werner
state is introduced. The mapping from infinite-dimensional Hilbert space to
Hilbert space of two qubits is described in Sec.~III. In Sec.~IV, we will
analyse the separability of the Werner state from two different points of view:
after and before mapping on two-qubit system. Sec.~V and Sec.~VI are
devoted to the nonlocality and squeezing of the Werner state. In
Sec.~VII, the coherent-state teleportation with the Werner state is
discussed. Finally, Sec.~VIII contains the conclusions.
\section{Continuous-variable Werner state}

A common resource of the quantum entanglement in CV information processing
is the two-mode squeezed vacuum state generated by means of spontaneous parametric
downconversion in the non-degenerate optical parametric amplifier (NOPA),
\begin{equation}\label{rhonopa}
{\rho}_{\rm NOPA}=(1-{\lambda}_{1}^2)\sum_{m,n=0}^{\infty}
{\lambda}_{1}^{m+n}|m,m\rangle\langle n,n|.
\end{equation}
Here $\lambda_{1}=\tanh r$, $r$ is the squeezing parameter,
and $|m,n\rangle=|m\rangle_{A}|n\rangle_{B}$ denotes the Fock
state of two modes $A$ and $B$. The NOPA state approaches the
maximally entangled EPR state \cite{EPR} in the strong squeezing limit
$r\to\infty$. In practice, the EPR state is well approximated by the NOPA
state if $r>2$. Recently, squeezing as large as $r\approx 2$ has been
achieved experimentally \cite{Kumar}.

A natural extension of the NOPA state to the Werner state for CV is based
on the following point of view: The factorized state $I_1 \otimes I_2/d^2$
in the mixture (\ref{werner}) is a tensor product of density matrices
$I_1/d$ and $I_2/d$ which can be identified with reduced density matrices
of the subsystems $1$ and $2$ when the whole system is in the maximally
entangled state $|\Psi\rangle$. Now, if the modes $A$ and $B$
are in the NOPA state, then each mode separately is in thermal state.
The thermal state of modes $A$ and $B$ can be expressed as follows,
\begin{equation}\label{rhot}
{\rho}_{\rm T}=(1-{\lambda}_{2}^{2})^{2}\sum_{m,n=0}^{\infty}
{\lambda}_{2}^{2(m+n)}|m\rangle\langle m|\otimes |n\rangle\langle n|,
\end{equation}
where $\lambda_{2}=\tanh s$ and   the mean number of
thermal photons in each mode reads $\langle n\rangle_{T}=\sinh^{2}(s)$.

It is thus natural to define the CV analogue of the Werner state
\cite{Werner89} as a mixture of the NOPA state (\ref{rhonopa}) and
factorized thermal state (\ref{rhot}),
\begin{equation}\label{rhow}
{\rho}_{\rm W}=p\,{\rho}_{\rm NOPA}+(1-p)\,{\rho}_{\rm T},\quad 0\leq
p\leq 1.
\end{equation}
The Werner states $\rho_W$ form a three-parametric family of states.
The simplest analogue of Werner state can be obtained assuming $r=s$.
In this case the Werner state and $d$-dimensional Werner state (\ref{werner})
become manifestly analogous in the strong squeezing limit when $\rho_W$
approaches a mixture of maximally entangled EPR state and maximally mixed
state in infinite-dimensional Hilbert space.
\section{Mapping on two-qubit system}

The simplest way in which one can study the separability and
nonlocality properties of the two-mode state (\ref{rhow}) is to map it
by means of {\em local operations} on
the two-qubit system for which separability and nonlocality
conditions are well known \cite{Peres96,Horodecki96}. In
what follows the qubits corresponding to modes A~and B are denoted
as $1$ and $2$, respectively.

We introduce the Hermitian ``spin one-half'' operators $S_j^\alpha$,
$\alpha=A,B$,
\begin{equation}\label{smatrix}
\begin{array}{c}
\displaystyle
S_{1}^{\alpha}+iS_2^\alpha
=2\sum_{m=0}^{\infty}|2m\rangle_{\alpha\alpha}\langle2m+1|,
\\[6mm]
\displaystyle
S_{3}^{\alpha}=\sum_{m=0}^{\infty}(-1)^m|m\rangle_{\alpha\alpha}\langle m|,
\end{array}
\end{equation}
which satisfy the Pauli matrix algebra
\begin{equation}\label{commut}
[S_i^{\alpha},S_j^{\beta}]=2i\varepsilon_{ijk}\delta_{\alpha\beta}
S_{k}^{\alpha}, \qquad (S_i^{\alpha})^{2}=1,
\end{equation}
where $\varepsilon_{ijk}$ is the totally antisymmetric tensor
with $\varepsilon_{123}=+1$ and $\delta_{\alpha\beta}$ is the
Kronecker symbol.

Let us for a moment restrict our attention to the mode $A$ and qubit $1$.
With the help of the  operators (\ref{smatrix})
one can assign the following qubit density matrix $\rho_{1}$
to the density matrix $\rho_{A}$,
\begin{eqnarray}\label{rho3}
\rho_{1}&=&\frac{1}{2}(I_{1}+{\bf S}^{A}\cdot\bm{\sigma}),
\end{eqnarray}
where
\begin{equation}\label{ssmatrix}
{\bf S}^{A}\cdot\bm{\sigma}=\sum_{i=1}^{3}\mbox{Tr}(\rho_{A}
S_{i}^{A})\sigma_{i},
\end{equation}
$\sigma_{i}$ are standard Pauli matrices and $I_{1}$ is the identity
operator on the Hilbert space of qubit $1$.

The transformation (\ref{rho3}) is physical, because it  can be,
at least in principle, performed in the lab.
Let us assume that the qubit is represented by
a two-level atom resonantly interacting with a single mode of electromagnetic
field. Suppose that the interaction is governed by the following Hamiltonian
\begin{equation}\label{H}
H_{\rm int}=i\hbar \Omega \left(|0\rangle\langle 1| \sqrt{n}
a^\dagger -|1\rangle\langle 0| a \sqrt{n}\right),
\end{equation}
where $a$ ($a^\dagger$) is annihilation (creation) operator of the mode
$A$ and $n=a^\dagger a$.
The Hamiltonian (\ref{H}) can be considered as a kind of nonlinear
Jaynes-Cummings model. The specific feature of $H_{\rm int}$ is that its
eigenvalues (the Rabi frequencies) are linearly proportional to the
number of photons $n$. If the two-level atom is initially in its ground
state $|0\rangle$ and if the interaction time $t$ is adjusted in such a
way that $\Omega t=\pi/2$ then the output state of the atom is exactly
given by Eq. (\ref{rho3}).  Although the Hamiltonian (\ref{H}) may be hard
to implement in practice, it provides a clear physical picture behind
the mathematical transformation (\ref{rho3}).

Formally, the transformation (\ref{rho3}) is a trace-preserving completely
positive (CP) map. Making use of the correspondence between CP maps and
positive semidefinite operators \cite{Fiurasek01}
we can express the transformation (\ref{rho3}) as follows,
\begin{eqnarray}\label{rho1}
\rho_{1}=\mbox{Tr}_{A}[\chi_{A1}\rho_{A}^{T}\otimes I_{1}],
\end{eqnarray}
where
\begin{eqnarray}\label{chiA1}
\chi_{A1}&=&\sum_{m=0}^{\infty}\sum_{k,l=0}^{1}
(|2m+k\rangle_{AA}\langle 2m+l|\otimes|k\rangle_{11}\langle l|\nonumber\\
\end{eqnarray}
is a positive-semidefinite  operator acting on the direct product
of the Hilbert spaces of the mode $A$ and of the qubit $1$;
$|1\rangle_{1}$ and $|0\rangle_{1}$ are basis states of qubit $1$;
and $T$ stands for the transposition. Mapping now the two-mode
density matrix (\ref{rhow}) to two-qubit density matrix
\begin{eqnarray}\label{rhoW'}
\rho_{\rm W}'=\mbox{Tr}_{AB}[\chi_{A1}\chi_{B2}\rho_{\rm
W}^{T}\otimes I_{1}\otimes I_{2}],
\end{eqnarray}
where the $\chi_{B2}$ is obtained from (\ref{chiA1}) by replacements
$A\to B$ and $1\to 2$, one gets
\begin{equation}\label{rhow2}
\rho_{\rm W}'=\frac{1}{4}(I_{1}\otimes I_{2}+{\bf S}^{A}\cdot
\bm{\sigma}\otimes I_{2}+I_{1}\otimes{\bf S}^{B}\cdot\bm{\sigma}
+\sum_{i,j=1}^{3}t_{ij}\sigma_{i}\otimes\sigma_{j}).
\end{equation}
The elements $t_{ij}=\mbox{Tr}(\rho_{\rm W}S_{i}^{A}S_{j}^{B})$
of the correlation tensor $T$ explicitly read
\begin{eqnarray}\label{corelt}
t_{11}&=&-t_{22}=\frac{2\lambda_{1}p}{1+\lambda_{1}^2},
\nonumber \\
t_{33}&=&p+(1-p)\left(\frac{1-\lambda_{2}^2}{1+\lambda_{2}^{2}}
\right)^2,\nonumber\\
t_{ij}&=&0,\quad i\ne j.
\end{eqnarray}
On calculating the matrices ${\bf S}^{A}\cdot\bm{\sigma}$
and ${\bf S}^{B}\cdot\bm{\sigma}$ and taking into account the
expressions (\ref{corelt}) one obtains after some algebra
\begin{widetext}
\begin{equation}\label{rhow2b}
{\rho}_{\rm W}'=
\left(
\begin{array}{cccc}
\displaystyle
\frac{p}{1+{\lambda}_{1}^{2}}+\frac{1-p}{(1+
{\lambda}_{2}^{2})^{2}} & \displaystyle 0 & \displaystyle 0 &
\displaystyle \frac{\lambda_1 p}{1+{\lambda}_{1}^{2}}\\
\displaystyle  0 &
\displaystyle \frac{\lambda_{2}^{2}(1-p)}{(1+{\lambda}_{2}^{2})^{2}} &
\displaystyle  0 &\displaystyle   0 \\
\displaystyle  0 & \displaystyle 0 &
 \displaystyle  \frac{\lambda_{2}^{2}(1-p)}{(1+{\lambda}_{2}^{2})^{2}} &
 \displaystyle  0 \\
\displaystyle  \frac{\lambda_1 p}{1+{\lambda}_{1}^{2}} &
\displaystyle  0 &\displaystyle   0 &
\displaystyle \frac{\lambda_{1}^{2}p}{1+{\lambda}_{1}^{2}}+
 \frac{\lambda_{2}^{4}(1-p)}{(1+{\lambda}_{2}^{2})^{2}}
\end{array}\right).
\end{equation}
\end{widetext}
Thus we  have mapped the state ${\rho}_{W}$ onto this two-qubit state.
Note that the transformation  (\ref{rhoW'}) is local; it is carried out
separately on each subsystem $(A,1)$ and $(B,2)$. The essential feature
of local unconditional transformations is that they cannot increase
the amount of entanglement or nonlocality present in any
bipartite state. This ensures that the properties of the state $\rho_W'$
reflect the properties of the original Werner state $\rho_W$.
If we find that the state $\rho_W'$ is entangled or nonlocal, then the
same holds true about the original state $\rho_W$.
\section{Separability}

According to the PH partial transposition criterion
\cite{Peres96,Horodecki96} the state (\ref{rhow2}) is
entangled iff its partial transpose
\begin{eqnarray}\label{partrho}
(\rho_{\rm W}')_{m\mu,n\nu}^{T_{1}}\equiv(\rho_{\rm
W}')_{n\mu,m\nu},
\end{eqnarray}
has some negative eigenvalue. Due to the specific structure of the
matrix (\ref{rhow2b}), it is easy to see that its partial transposition
has negative eigenvalue if the off-diagonal elements of $\rho_W'$ are
larger than the central diagonal elements,
\begin{equation}
\frac{\lambda_{1}p}{1+{\lambda}_{1}^{2}}>
\frac{\lambda_{2}^{2}(1-p)}{(1+{\lambda}_{2}^{2})^{2}}.
\end{equation}
It is instructive to rewrite this condition as an inequality for the
probability $p$. After some algebra, one finds that the state (\ref{rhow2})
is entangled iff
\begin{eqnarray}\label{entangl2}
p>\frac{1}{1+2\frac{\tanh(2r)}{\tanh^{2}(2s)}},
\end{eqnarray}
where we have used the relations $\lambda_1=\tanh r$ and
$\lambda_2=\tanh s$. For direct analogue of the Werner state with $r=s$,
and in the large squeezing limit, the state (\ref{rhow2b}) and hence
also the state (\ref{rhow}) are entangled if $p>\frac{1}{3}$ as
in the case of two-qubit Werner state \cite{Dur00}.

Surprisingly, the PH criterion can also be applied
directly to the two-mode state (\ref{rhow}) for which it is only
sufficient condition for entanglement. The partially
transposed matrix $\rho_{\rm W}^{T_{A}}$ has
a block diagonal form with $1\times 1$ blocks in one-dimensional
subspaces spanned by vectors $\{|m,m\rangle\}$,
$m=0,1,...$ and $2\times 2$ blocks in two-dimensional subspaces spanned
by vectors $\{|m,n\rangle, |n,m\rangle, m\ne n\}$, $m,n=0,1,...$.
Consequently,
the eigenvalues of the partially transposed matrix $\rho_{\rm W}^{T_{A}}$
can easily be calculated  as roots of quadratic equations and read
\begin{eqnarray}\label{eigen}
x^{(l)}&=&p(1-\lambda_{1}^{2})\lambda_{1}^{2l}+
(1-p)(1-\lambda_{2}^{2})^{2}\lambda_{2}^{4l},\nonumber\\[2mm]
x_{1,2}^{(mn)}&=&(1-p)(1-\lambda_{2}^{2})^{2}\lambda_{2}^{2(m+n)}\pm
p(1-\lambda_{1}^{2})\lambda_{1}^{m+n},
\nonumber \\
\end{eqnarray}
where $l=0,1,...$ and $m\ne n=0,1,...$. According to the above
mentioned separability criterion \cite{Peres96} the state
(\ref{rhow}) is entangled if there are such $m$, $n$ for which
$x_{2}^{(mn)}<0$. If $\lambda_{2}=0$ then $x_{2}^{(mn)}<0$
for all $p>0$ and the state (\ref{rhow}) is always entangled.
If $\lambda_{2}\ne 0$
then the inseparability condition $x_{2}^{(mn)}<0$ is equivalent
with the inequality
\begin{eqnarray}\label{entanglp}
p>\frac{(1-\lambda_{2}^{2})^{2}}{(1-\lambda_{2}^{2})^{2}+
(1-\lambda_{1}^{2})\left(\frac{\lambda_{1}}{\lambda_{2}^{2}}
\right)^{m+n}}\equiv p_{m+n}.
\end{eqnarray}
Three different cases must be considered in dependence on the
value of the ratio $q=\lambda_1/\lambda_2^2$.

(i) If $q>1$ then the factor $q^{m+n}$ in the denominator of Eq.
(\ref{entanglp}) increases with increasing $m+n$ and consequently
the right-hand side (R.H.S.) of (\ref{entanglp}) decreases attaining
zero value in the limit $m+n\to \infty$.
Hence, the state (\ref{rhow}) is entangled for any
$p>0$. In particular, a direct analogue of the Werner state ($r=s$) is
entangled for every $p>0$.

(ii) If $q=1$ then also $q^{m+n}=1$. From that it follows that the
inequality (\ref{entanglp}) is independent on $m$ and $n$ and the
state (\ref{rhow}) is entangled if
\begin{eqnarray}\label{ineq1}
p>\frac{1-\lambda_{1}}{2}=\frac{1-\tanh r}{2}.
\end{eqnarray}
(iii) If $q<1$ then the R.H.S. of inequality
(\ref{entanglp}) increases with growing $m+n$. Since the R.H.S.
attains its minimum value for $m+n=1$ the state (\ref{rhow}) is
entangled if
\begin{equation}\label{ineq2}
p>\frac{(1-\lambda_{2}^{2})^{2}}{(1-\lambda_{2}^{2})^{2}+
(1-\lambda_{1}^{2})\frac{\lambda_{1}}{\lambda_{2}^{2}}}=p_1
\end{equation}
or equivalently,
\begin{equation}
p>\frac{1}{1+\frac{\cosh^{4}(s)}{\cosh^{2}(r)}\frac{\tanh
r}{\tanh^{2}(s)}}.
\end{equation}
The partial transposition criterion applied to the original
state $\rho_W$ is stronger than that applied to $\rho_W'$, because any
local transformation (\ref{rhoW'}) preserves the positivity of the partial
transpose. For instance,  if $\tanh r>\tanh^{2}(s)$ then the
entangled states (\ref{rhow}) for which
\begin{eqnarray}\label{ineq3}
\frac{\tanh^{2}(2s)}{\tanh^{2}(2s)+2\tanh(2r)}\geq p>0
\end{eqnarray}
are mapped on the separable two-qubit states (\ref{rhow2}).

The region of Werner state inseparability is depicted in Fig.~1.
We can see that the Werner state is entangled almost for every $p$ if
the squeezing is sufficiently large. We emphasize again that the negative
partial transpose is only sufficient condition on the entanglement and there
may exist PPT entangled Werner states. One may even ask whether there exist
any nontrivial separable CV Werner states (\ref{rhow}).
Although we do not have the sufficient separability condition for
generic bipartite CV states at present, it is possible to find conditions
under which the Werner state is separable, i.e. it can be written as a
convex combination of product states.

The state (\ref{rhow}) can be rewritten in the following form
\begin{eqnarray}\label{rhowsep}
\rho_{\rm W}=\sum_{m=0}^\infty P_m |mm\rangle\langle mm|
+\sum_{m\ne n=0}^{\infty}\rho^{mn},
\end{eqnarray}
where
\begin{equation}
P_m=p(1-\lambda_1^2)^2 \lambda_{1}^{4m}+(1-p)(1-\lambda_2^2)^{2}
(1-\lambda_2^4)\lambda_{2}^{8m},
\end{equation}
and $\rho^{mn}$ are matrices in four-dimensional Hilbert subspaces
spanned by the basis vectors $\{$$|mm\rangle,$ $|mn\rangle,$ $|nm\rangle,$
$|nn\rangle\}$,
\begin{equation}\label{rhomn}
\rho^{mn}=\frac{1}{2}\left(
\begin{array}{cccc}
\alpha_{mn} & 0 & 0 & \beta_{mn} \\
0 & \gamma_{mn} & 0 & 0 \\
0 & 0 & \gamma_{mn} & 0 \\
\beta_{mn} & 0 & 0 & \alpha_{mn}
\end{array}
\right),
\end{equation}
 where
\begin{eqnarray}\label{koef}
\alpha_{mn}&=&p(1-\lambda_{1}^{2})^{2}\lambda_{1}^{2(m+n)}
\nonumber\\
&&+{(1-p)(1-\lambda_{2}^{2})^{2}(1-\lambda_{2}^{4})\lambda_{2}^
{4(m+n)}},\nonumber\\
\beta_{mn}&=&p(1-\lambda_{1}^{2})\lambda_{1}^{m+n},
\nonumber\\
\gamma_{mn}&=&(1-p)(1-\lambda_{2}^{2})^{2}\lambda_{2}^{2(m+n)}.
\end{eqnarray}
Obviously, if all $\rho^{mn}$ are positively semidefinite
separable matrices, then $\rho_{\rm W}$ is separable.
From the matrix form (\ref{rhomn}) of $\rho^{mn}$ one easily obtains
the positivity condition
$\alpha_{mn}\geq\beta_{mn}$ and the separability condition
$\gamma_{mn}\geq\beta_{mn}$. Consequently, the Werner state is
separable if both these inequalities are satisfied for all $m\ne n$.
The second condition is identical with the necessary condition
on separability of the Werner state $p \leq p_{m+n}$, where $p_{m+n}$ is
given in the inequality (\ref{entanglp}). Further constraint on $p$
follows from the positivity condition $\alpha_{mn}\geq\beta_{mn}$,
\begin{equation}\label{entanglpb}
p\leq\frac{1}{1+\frac{(1-\lambda_{1}^{2})}{(1-\lambda_{2}^{2})^{2}
(1-\lambda_{2}^{4})}\left[\left(\frac{\lambda_{1}}{\lambda_{2}^{4}}
\right)^{m+n}-(1-\lambda_{1}^{2})\left(\frac{\lambda_{1}^{2}}
{\lambda_{2}^{4}}\right)^{m+n}\right]},
\end{equation}
where $m\ne n=0,1,...$.
Similarly as in the case of entanglement we have to distinguish
three regions in dependence on the value of the ration
$\tilde{q}=\lambda_1/\lambda_2^4$:

(i) If $\tilde{q}>1$ then the R.H.S. of the
inequality (\ref{entanglpb}) goes to zero in the limit $m+n\to \infty$.
Hence, the state (\ref{rhow}) is separable only for $p=0$.

(ii) If $\tilde{q}=1$ then the expression in
square brackets in the R.H.S. of the inequality (\ref{entanglpb}) attains
its maximum equal to unity in the limit $m+n\to \infty$ and the Werner
state can be separable only if
\begin{equation}\label{sepineq1}
p\leq\frac{(1-\lambda_{2}^{2})^{2}}{2(1-\lambda_{2}^{2}+\lambda_{2}^
{4})}.
\end{equation}
In this case the condition $p\leq p_1$ is weaker than the inequality
(\ref{sepineq1}) which is thus sufficient condition for
separability of the Werner state.

(iii) If $\tilde{q}<1$ then the R.H.S. of the
inequality (\ref{entanglpb}) attains its minimum for $m+n=1$
and the state (\ref{rhow}) can be separable only if
\begin{eqnarray}\label{sepineq2}
p\leq\frac{1}{1+\frac{(1-\lambda_{1}^{2})}{(1-\lambda_{2}^{2})^{2}
(1-\lambda_{2}^{4})}\left[\left(\frac{\lambda_{1}}{\lambda_{2}^{4}}
\right)-(1-\lambda_{1}^{2})\left(\frac{\lambda_{1}^{2}}
{\lambda_{2}^{4}}\right)\right]}.
\end{eqnarray}
Since in this case the inequality (\ref{sepineq2}) is stronger than
the condition $p\leq p_1$, the inequality (\ref{sepineq2})
is sufficient condition for separability of the Werner state.
\section{Nonlocality}

Due to the commutation rules (\ref{commut}) the nonlocality of the
Werner state (\ref{rhow}) can be investigated employing the standard
two-qubit CHSH Bell inequalities in which the Pauli matrices are
replaced with the single-mode operators (\ref{smatrix}),
\begin{eqnarray}\label{Bell}
2\ge|\langle({\bf a}\cdot{\bf S}^{A})({\bf b}\cdot{\bf
S}^{B})\rangle+\langle({\bf a}'\cdot{\bf S}^{A})({\bf b}\cdot{\bf
S}^{B})\rangle\nonumber\\
+\langle({\bf a}\cdot{\bf S}^{A})({\bf b}'\cdot
{\bf S}^{B})\rangle-\langle({\bf a}'\cdot{\bf S}^{A})({\bf
b}'\cdot{\bf
S}^{B})\rangle| ,
\end{eqnarray}
where ${\bf a}$, ${\bf a}'$, ${\bf b}$, ${\bf b}'$ are real
three-dimensional unit vectors and the angle brackets denote the
averaging over the density matrix $\rho_{W}$. It is instructive to
formulate this approach in terms of the mapping introduced in Sec. III.
We map the Werner state $\rho_W$ onto the state of two qubits $\rho_W'$
and then we analyze the nonlocality of the state $\rho_W'$ characterized
by the correlation tensor $T$ whose elements are given by Eq.~(\ref{corelt}).

Now, according to the Horodecki criterion \cite{Horodecki95}, if the sum of
the two largest eigenvalues of the matrix $U=T^{T}T$ is greater than unity
then the state (\ref{rhow}) violates the inequalities (\ref{Bell}) for
some choice of vectors ${\bf a}$, ${\bf a}'$, ${\bf b}$, ${\bf b}'$.
The matrix $U$ has two-fold eigenvalue $t_{11}^{2}$ and
single eigenvalue $t_{33}^{2}$. It can be shown that the
inequality $t_{11}^{2}\leq t_{33}^{2}$ is satisfied for any
$\lambda_1$, $\lambda_2$  and $p$.
Thus the maximal Bell factor is given by
\begin{equation}
B_{\mbox{max}}=2\sqrt{t^{2}_{11}+ t^{2}_{33}}.
\end{equation}
Hence, the Bell inequality (\ref{Bell}) is violated if
$t_{11}^{2}+t_{33}^{2}>1$. Substituting here from the formulas
(\ref{corelt}) one obtains after some algebra that the state
(\ref{rhow}) violates the Bell inequalities (\ref{Bell}) if
\begin{equation}\label{pnonloc}
p>\frac{a(a-1)+\sqrt{a(a-ab^{2}+2b^{2})}}{a^{2}+b^{2}},
\end{equation}
where $a=\tanh^{2}(2s)$ and $b=\tanh(2r)$.
The region of nonlocality of the state (\ref{rhow2}) is depicted
in Fig.~2. In the large squeezing limit the direct
analogue of the Werner state (\ref{rhow}) with $r=s$ is nonlocal
if $p>\frac{1}{\sqrt{2}}$ as in the case of two-qubit Werner state. However,
it was found in the previous Section, that the original state $\rho_W$ in
infinite-dimensional Hilbert space may be entangled even if the two-qubit
state $\rho_W'$ is separable. We can conjecture that the nonlocality has a similar behavior
and that the state $\rho_W$ may violate some Bell inequalities
although the state $\rho_W'$ admits local realistic description.
\section{Squeezing}

Apart from entanglement and nonlocality, the measure of a nonclassicality
of the Werner state (\ref{rhow}) can be judged by means of squeezing.
Since there is not preferred direction in the phase-space distribution
of the thermal component (\ref{rhot}) of the Werner state (\ref{rhow})
one can expect that the state attains its maximum squeezing in the same
quadrature as the NOPA state (\ref{rhonopa}), i.e. in the quadrature
$x_{A}-x_{B}$ (where $x_{A}$, $x_{B}$ are "position" quadratures of modes
$A$ and $B$, respectively).
Calculating the variance $\langle[\Delta(x_{A}-x_{B})]^{2}\rangle$ in
the Werner state (\ref{rhow})  and employing the squeezing condition
$\langle[\Delta(x_{A}-x_{B})]^{2}\rangle<1$ one finds, that the Werner
state is squeezed if
\begin{eqnarray}\label{squeezing}
p>\frac{1}{1+\frac{2\lambda_{1}(1-\lambda_{1}^{2})}{(1+\lambda_{1})
(1+3\lambda_{2}^{2})}}=\frac{1}{1+\frac{1-e^{-2r}}{1+4\langle
n\rangle_{T}}}.
\end{eqnarray}
Interestingly, since for $r=s$ and in the large squeezing limit the R.H.S.
of the inequality goes to unity, we arrive at the family of Werner
states which are never squeezed. This is the counterintuitive example of the
Werner states which are not squeezed, however, they can be entangled or even
nonlocal.
\section{Teleportation}

An interesting application of two-qubit Werner state arises in the quantum
teleportation \cite{Bennett93,Horodecki99b}. By analogy, the proposed
Werner state (\ref{rhow}) can be utilized in BK scheme of the CV
coherent-state teleportation \cite{Braunstein98}.
In this scheme, two-mode squeezed vacuum state is shared between
Alice and Bob. The Wigner function $W_{in}(x_{1},p_{1})$ of the
input Alice's state and the Wigner function $W_{out}(x_{2},p_{2})$ of
the Bob's output state are related by the convolution \cite{Chizhov}
\begin{eqnarray}
W_{out}(x_{2},p_{2})&=&\frac{1}{4}\int_{-\infty}^{\infty}
K_{AB}(x_{2}-x_{1},p_{2}-p_{1})\times\nonumber\\
& &W_{in}(x_{1},p_{1})dx_{1}dp_{1}.
\end{eqnarray}
The kernel function $K_{AB}(x_{-},p_{+})$ reads
\begin{equation}
K_{AB}(x_{-},p_{+})=\int_{-\infty}^{\infty}{\cal W}_{AB}
(-x_{-},x_{+},p_{-},p_{+})dx_{+}dp_{-},
\end{equation}
where $x_{\pm}=x_{A}\pm x_{B}$ and $p_{\pm}=p_{A}\pm p_{B}$, and
$W_{AB}(x_{A},p_{A},x_{B},p_{B})={\cal W}_{AB}(x_{-},x_{+},p_{-},p_{+})$
is Wigner function of the state shared by Alice and Bob (quantum
channel). The fidelity between Alice's and Bob's states can be calculated
as follows,
\begin{eqnarray}
F&=&\frac{\pi}{2}\int_{-\infty}^{\infty}\int_{-\infty}^{\infty}
W_{in}(x_{2},p_{2})K_{AB}(x_{2}-x_{1},p_{2}-p_{1})\times\nonumber\\
& &W_{in}(x_{1},p_{1})
dx_{1}dp_{1}dx_{2}dp_{2}.
\end{eqnarray}
The BK scheme is designed in such a way, that the fidelity is invariant
under displacement transformations. In particular, all coherent states are
teleported with the same fidelity. If the quantum channel is in NOPA state
then this fidelity reads
\begin{equation}
F_{NOPA}=\frac{1}{1+e^{-2r}}.
\end{equation}
This exceeds the maximal classical value $F=1/2$ for every $r>0$.
Now we consider the Werner state (\ref{rhow}) in symmetric form with $r=s$.
One can find that the fidelity of teleportation in standard BK scheme
utilizing such a Werner state is changed as follows
\begin{equation}\label{wfidel}
F_{W}=pF_{NOPA}+(1-p)\frac{1}{d},
\end{equation}
where $d=2/(1-\lambda_{1}^{2})=2\cosh^{2}r$. The dependence of $F_W$ on
probability $p$ and squeezing parameter $r$ is depicted in Fig.~3.

In the limit of large squeezing, the fidelity (\ref{wfidel}) approaches
a value $F_{W}\approx p$. In order to teleport the coherent state
with fidelity $F_{W}>1/2$, we need to employ CV Werner state with $p>1/2$.
Note that for every $p\not= 0$, this Werner state is entangled (recall
that we assume $r=s$ here). This should be contrasted with results
obtained for teleportation with $d$-dimensional Werner state,
where if the shared Werner state is entangled then it is useful for
quantum teleportation \cite{Badziag00}. In our case, some of the entangled
Werner states in infinite-dimensional Hilbert space are not useful for BK
teleportation protocol. It is an open question whether the BK scheme can
be modified in such a way that the coherent states would be teleported
with fidelity higher than $1/2$ even when using entangled Werner states
with $p<1/2$.
\section{Conclusions}

A natural extension of the Werner state into CV systems is
presented and separability, nonlocality and squeezing of this state is
analyzed. In certain sense, the CV  Werner state can be considered
as a counterpart of the two-qubit Werner state. This relationship is
established by the mapping (\ref{chiA1}). On the other hand, some features
of the CV Werner state correspond to those of $d$-dimensional Werner states
when $d\to \infty$. For instance, in the simplest case when $r=s$, the
CV Werner state (\ref{rhow}) is entangled for any $p>0$.
Since $d$-dimensional Werner state is entangled when $p>1/(1+d)$,
the above behavior of CV Werner state corresponds to the limit
$d\to \infty$.
\acknowledgments

This work was supported by the EU grant under QIPC project
IST-1999-13071 (QUICOV) and project LN00A015 of the Czech Ministry of
Education.

\newpage
\thispagestyle{empty}
\begin{figure}
\caption{Minimal probability $p_{\mbox{min}}$ characterizing the
entanglement of the CV Werner state as a function of the squeezing parameter
$r$ and the thermal noise parameter $s$. The state $\rho_{\rm W}$ is
entangled if $p>p_{\mbox{min}}$.}
\end{figure}
\begin{figure}
\caption{Maximal probability $p_{\mbox{max}}$ characterizing the separability
of the CV Werner state as a function of the squeezing parameter $r$ and the
thermal noise parameter $s$. The state $\rho_{\rm W}$ is separable if
$p\leq p_{\mbox{max}}$.}
\end{figure}
\begin{figure}
\caption{Minimal probability $p_{\mbox{min}}$ characterizing the nonlocality
of the CV Werner state as a function of the squeezing parameter $r$ and
the thermal noise parameter $s$. The state $\rho_{\rm W}$ is nonlocal if
$p>p_{\mbox{min}}$.}
\end{figure}
\begin{figure}
\caption{The dependence of the fidelity $F_{W}$ in standard BK scheme employing
shared CV Werner state on the squeezing parameter $r=s$ and the probability
of the NOPA state $p$.}
\end{figure}

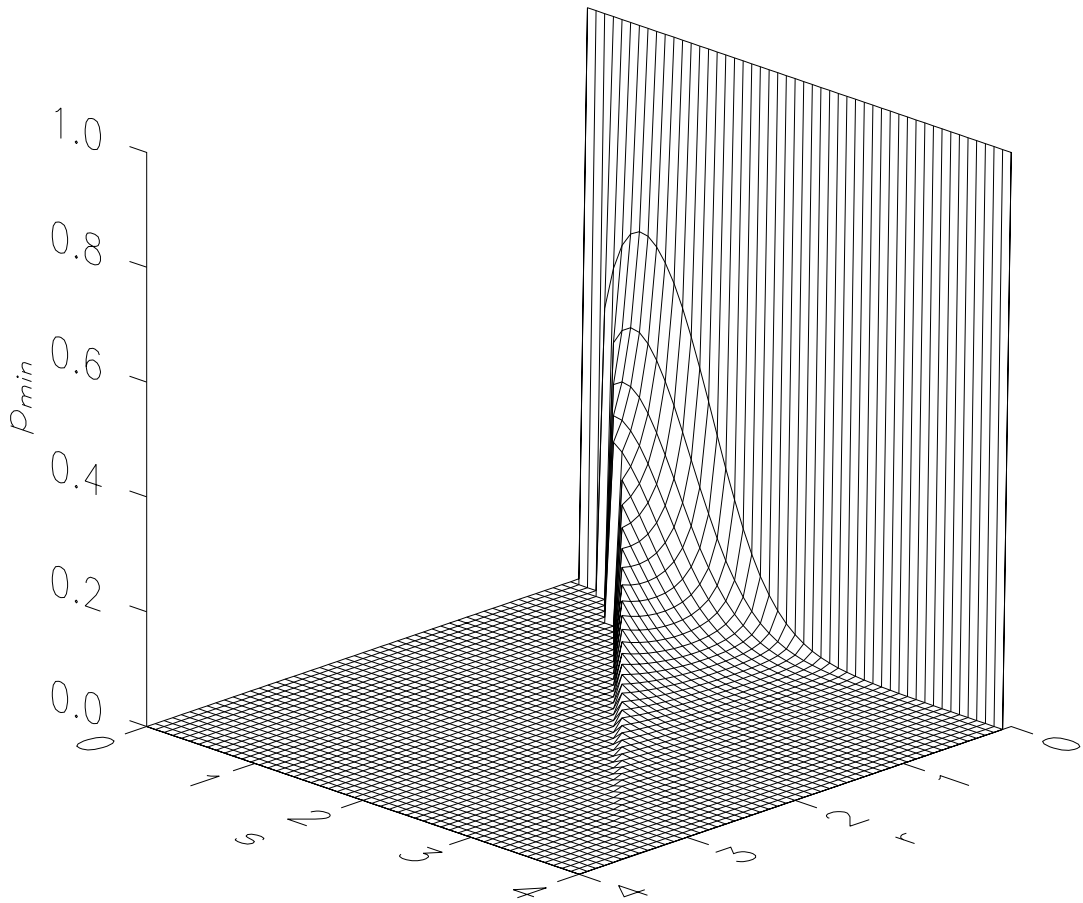
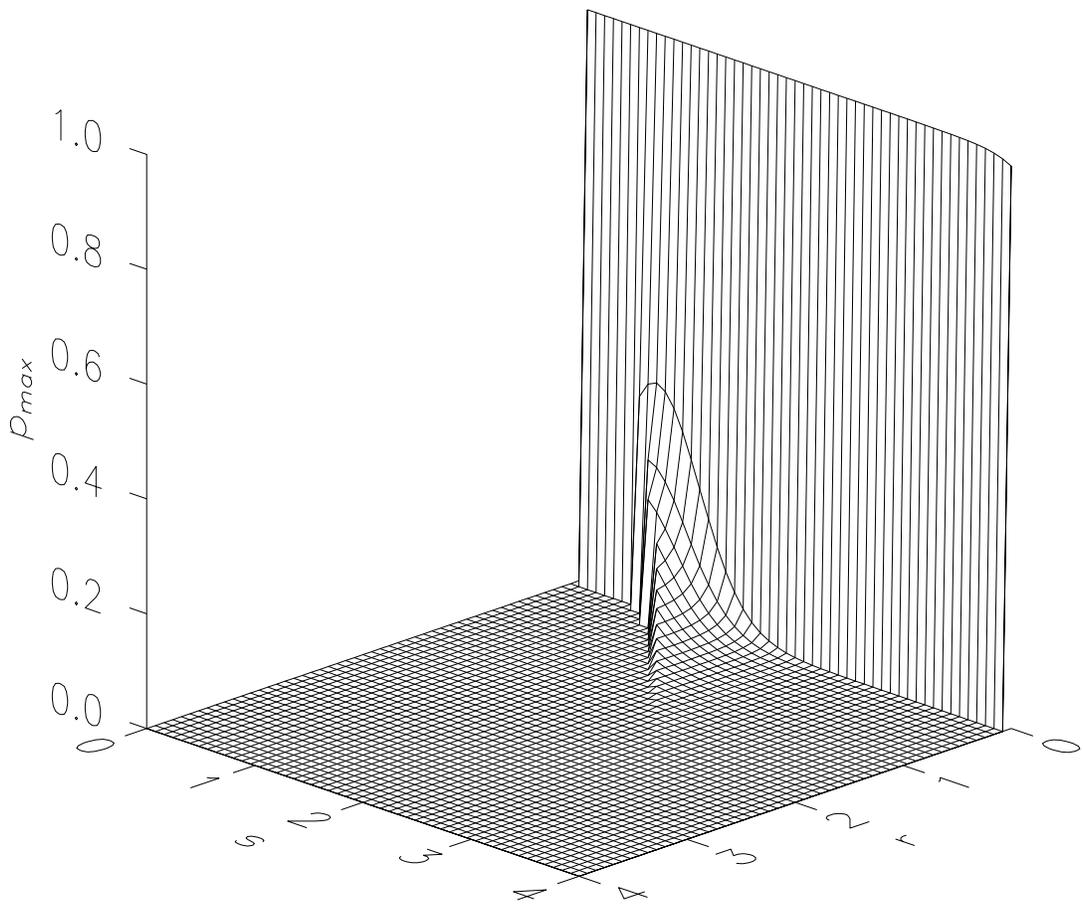
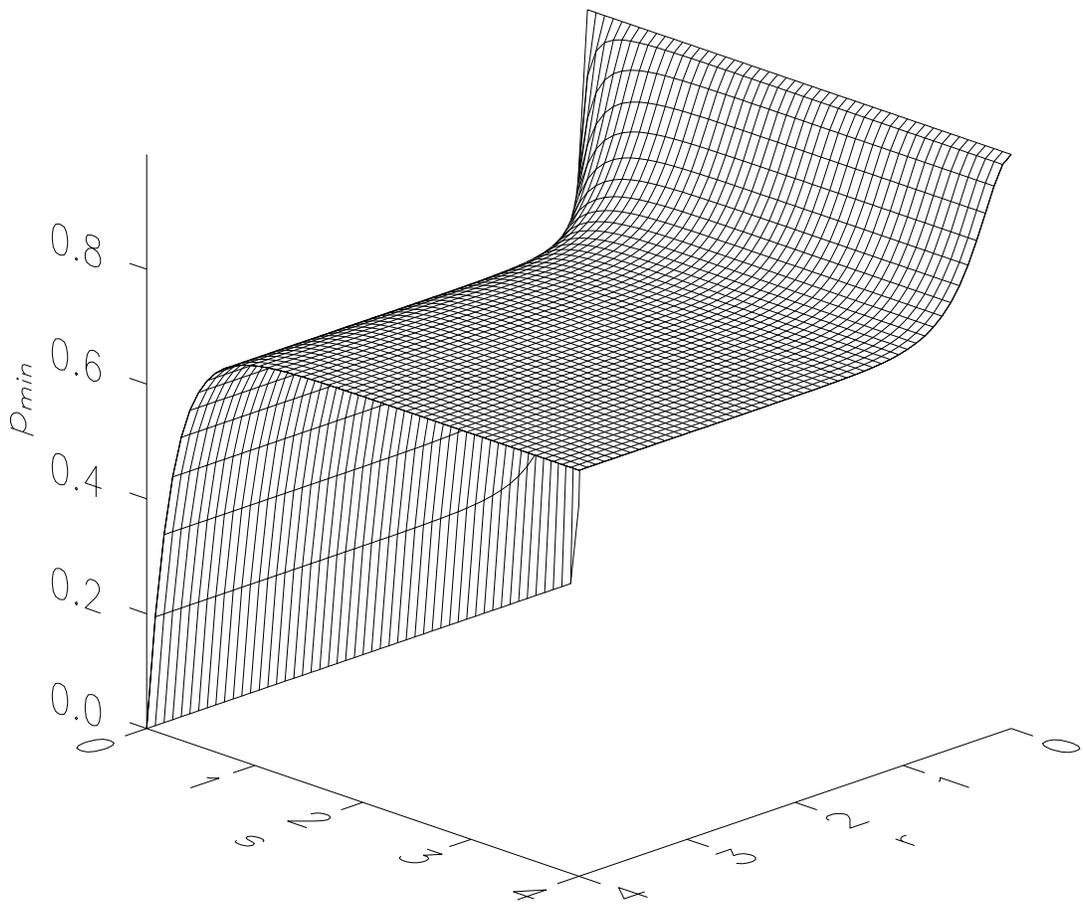
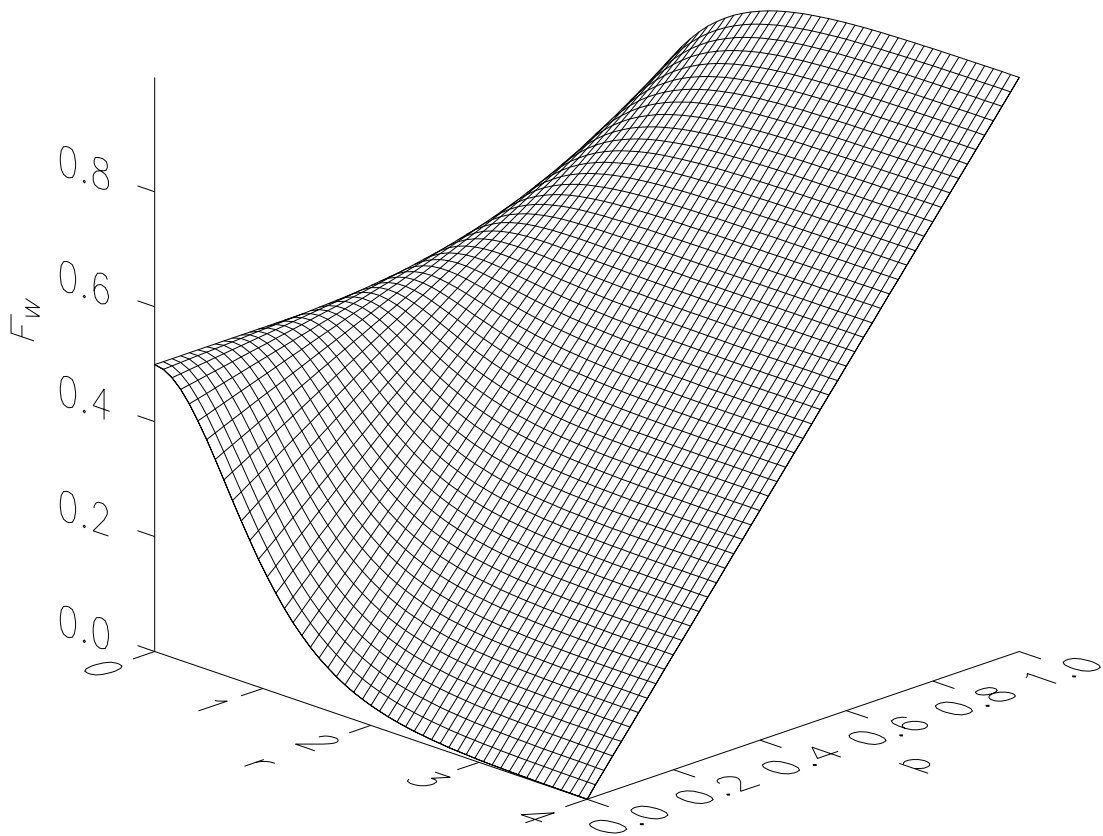
\end{document}